\documentclass[letterpaper]{article} % DO NOT CHANGE THIS
\usepackage{aaai24}  % DO NOT CHANGE THIS
\usepackage{times}  % DO NOT CHANGE THIS
\usepackage{helvet}  % DO NOT CHANGE THIS
\usepackage{courier}  % DO NOT CHANGE THIS
\usepackage[hyphens]{url}  % DO NOT CHANGE THIS
\usepackage{graphicx} % DO NOT CHANGE THIS
\usepackage{natbib}  % DO NOT CHANGE THIS AND DO NOT ADD ANY OPTIONS TO IT
\usepackage{caption} % DO NOT CHANGE THIS AND DO NOT ADD ANY OPTIONS TO IT
\usepackage{algorithm}
\usepackage{algorithmic}
\usepackage{amsmath}
\usepackage{amssymb}
\usepackage{soul}
\usepackage{bm}

\usepackage{newfloat}
\usepackage{listings}
\DeclareCaptionStyle{ruled}{labelfont=normalfont,labelsep=colon,strut=off} % DO NOT CHANGE THIS
\floatstyle{ruled}
\newfloat{listing}{tb}{lst}{}
\floatname{listing}{Listing}
\title{NeBLa: Neural Beer-Lambert for 3D Reconstruction of Oral Structures from Panoramic Radiographs}
\author{
    Sihwa Park\thanks{Contact Email: sihwapark@korea.ac.kr}\textsuperscript{\rm 1},
    Seongjun Kim\textsuperscript{\rm 1},
    Doeyoung Kwon\textsuperscript{\rm 1},
    Yohan Jang\textsuperscript{\rm 1},
    In-Seok Song\textsuperscript{\rm 2},
    Seung Jun Baek\thanks{Corresponding author.}\textsuperscript{\rm 1}
}
\affiliations{
    \textsuperscript{\rm 1}Korea University\\
    \textsuperscript{\rm 2}Korea University Anam Hospital
}

\begin{document}

\maketitle

\begin{abstract}
Panoramic radiography (Panoramic X-ray, PX) is a widely used imaging modality for dental examination. However, PX only provides a flattened 2D image, lacking in a 3D view of the oral structure. In this paper, we propose NeBLa (Neural Beer-Lambert) to estimate 3D oral structures from real-world PX. NeBLa tackles full 3D reconstruction for varying subjects (patients) where each reconstruction is based only on a single panoramic image. We create an intermediate representation called simulated PX (SimPX) from 3D Cone-beam computed tomography (CBCT) data based on the Beer-Lambert law of X-ray rendering and rotational principles of PX imaging. SimPX aims at not only truthfully simulating PX, but also facilitates the reverting process back to 3D data. We propose a novel neural model based on ray tracing which exploits both global and local input features to convert SimPX to 3D output. At inference, a real PX image is translated to a SimPX-style image with semantic regularization, and the translated image is processed by generation module to produce high-quality outputs. Experiments show that NeBLa outperforms prior state-of-the-art in reconstruction tasks both quantitatively and qualitatively. Unlike prior methods, NeBLa does not require any prior information such as the shape of dental arches, nor the matched PX-CBCT dataset for training, which is difficult to obtain in clinical practice. Our code is available at https://github.com/sihwa-park/nebla.
\end{abstract}

\section{Introduction}
Panoramic radiography (Panoramic X-ray, or PX) is a standard imaging technique for examining and diagnosing dental conditions. However, PX only provides flattened 2D images,  making it difficult to visualize the complete oral structure despite its panoramic nature. Since PX is a widely used imaging modality, there is a growing interest in estimating 3D oral structures from PX data. The estimated 3D representation offers a wide range of applications including 3D visualizations for patient education \cite{liang2020x2teeth}. Also, the generated 3D data can be integrated with virtual or augmented reality, e.g., in virtual surgical simulations for training physicians \cite{pohlenz2010virtual,li2021current}. Cone-beam computed tomography (CBCT) is another dental imaging technique which provides a comprehensive 3D view of oral and maxillofacial structures. CBCT enables accurate 3D imaging of hard tissues, but it is more expensive \cite{lopes2016comparison} and exposes patients to a greater dose of radiation compared to PX \cite{brooks2009cbct}.

Previous studies \cite{liang2020x2teeth, song2021oral} used conventional encoder-decoder models for 3D reconstruction from PX data. Those models are trained with \emph{synthetic} PX data generated from CBCT, instead of real-world PX. This is because it is difficult to obtain paired PX-CBCT data from the same patient. A number of problems arise from this approach. Firstly, the synthesis of PX typically involves flattening out layers of CBCT data \cite{yun2019automatic}, during which spatial information can be lost. Secondly, the encoder-decoder models are oblivious to how input PX are generated, and simply rely on supervision from synthetic data \cite{song2021oral}. Thirdly, to compensate for lost spatial information, the reconstruction model requires additional information such as the shape of patients' dental arches \cite{song2021oral}. To overcome these problems, we propose that the generation of synthetic PX and model architecture be closely intertwined. The PX synthesis should capture spatial information of CBCT which is based on the core principle of PX imaging. The model should be aware of the process of synthesizing 2D PX from 3D CBCT, so that the model can readily revert the process to reconstruct 3D data. Finally, one should narrow the domain gap between synthetic and real PX, so that CBCT can be accurately estimated from real PX.

Based on these guidelines, we propose a novel architecture named \emph{NeBLa} (\textit{\textbf{Ne}}ural \textit{\textbf{B}}eer-\textit{\textbf{La}}mbert) for the reconstruction of 3D oral structures from real-world PX images. To overcome the modality gap between real-world PX and CBCT, we propose an intermediate representation called simulated PX (SimPX). SimPX images are synthesized from CBCT by incorporating the imaging principles of PX and the Beer-Lambert law \cite{max1995optical} to render realistic PX images. Next, we propose a neural model for density estimation from SimPX inspired by NeRF \cite{mildenhall2020nerf}. However, unlike original NeRF which generates novel views of a single object, NeBLa estimates 3D structures for multiple objects (patients) using input PX image as a condition. Our model is designed to capture both local and global features of SimPX to an effective estimation of object-dependent densities. At the inference step, translation module is applied to convert the style from real PX to SimPX, where the model is trained for unpaired domain translation \cite{zhu2017unpaired} with semantic regularization to well-preserve the teeth region. Overall, the designs of  PX synthesis and model architecture of NeBLa are tightly coupled for enhanced reconstruction. Experiments show that NeBLa significantly outperforms state-of-the-art baselines in both quantitative and qualitative evaluation. 

Our contributions are summarized as follows: (1) We introduce a novel architecture based on ray tracing which reconstructs 3D oral structure from \emph{real-world PX}. To our knowledge, NeBLa is the first attempt to accomplish the task \emph{without any prior information} such as the shape of patients' dental arches; (2) We propose a new method for simulated PX based on PX imaging principles and the Beer-Lambert Law without the need for matching CBCT and PX datasets during training; (3) Experiments show that NeBLa can generate oral structures of improved quality, and is more robust to variations in real-world PX, compared to previous state-of-the-art methods.

\section{Related Work}
\subsection{2D-to-3D Reconstruction}
In recent years, there has been growing interest in methods for generating 3D models from 2D images, ranging from explicit techniques such as voxel-, point-, and mesh-based methods, to implicit function-based techniques. Voxel-based approaches \cite{brock2016generative,choy20163dr2n2,riegler2017octnet} have traditionally been a popular choice, however, the model size may grow exponentially with the target resolution of 3D models. Point-based methods \cite{fan2017point,wu2019pointconv} represent 3D objects as point clouds which can generate arbitrary-shaped 3D models with low computational complexity, but rely on point cloud datasets which may not be always available. Mesh-based approaches \cite{ranjan2018generating,wang2018pixel2mesh} generate 3D mesh models which represent 3D space with vertices and faces, allowing a detailed representation of 3D objects. However, generating high-quality meshes can be challenging due to surface irregularities or self-intersections. Recently, methods based on implicit functions  \cite{park2019deepsdf,mescheder2019occupancy,mildenhall2020nerf} have gained attention. Those methods utilize neural networks to represent object surfaces as continuous functions. NeRF \cite{mildenhall2020nerf} is a novel approach for 3D modeling, and directly estimates the radiance field of a scene, allowing for high-quality rendering of photorealistic 3D objects and scenes from arbitrary viewpoints.
%One of its variants is pixelNeRF \cite{yu2021pixelnerf} which can be trained with a sparse set of views via image-conditioning aligned with the input position vector. Recent research has investigated the limitations of NeRF such as the need for large training datasets and slow training time \cite{yu2021pixelnerf,yu2021plenoctrees,fridovich2022plenoxels}, and has extended its applicability to generative or language models \cite{schwarz2020graf,wang2022clip}.
NeBLa leverages the implicit function approach by NeRF for density estimation. The key difference is that, NeBLa reconstructs 3D structures of \emph{multiple} objects, each from a panoramic view to each object; NeRF performs 3D modeling of a \emph{single} object from multiple views to it.

\subsection{3D Reconstruction from X-ray}
Recent research on reconstructing 3D CT data or 3D volumes from 2D X-ray images has yielded promising results \cite{kasten2020end,henzler2018single}. CNN-based generative models are commonly used for 3D generation. For example, encoder-decoder networks which take an X-ray image as input and generate 3D structures are considered in \cite{henzler2018single,kasten2020end,liang2020x2teeth}. X2Teeth \cite{liang2020x2teeth} reconstructs individual teeth in 3D shapes using segmentation of the input PX image. GAN-based models have been explored, such as X2CT-GAN \cite{ying2019x2ct} and Oral-3D \cite{song2021oral} which reconstruct 3D thorax and oral structures, respectively. Oral-3D addresses a similar task to ours. However, it requires \emph{prior information} such as dental arches to reconstruct the oral structure. This is because their model uses synthesized PX images flattened from CBCT, and is trained to generate 3D flattened structures before making the final oral structure which needs a dental arch. One recent approach is NAF \cite{zha2022naf} which leverages NeRF \cite{mildenhall2020nerf} to reconstruct 3D CBCT data from multiple projections of a single object. NAF requires CBCT images projected from various angles, whereas NeBLa only uses a single PX image during the training for each object.

\section{Method}
\subsection{Overview}
Our objective is to reconstruct 3D oral structures from 2D X-ray images. The 2D images are panoramic radiographs (Panoramic X-ray or PX) whereas 3D data is obtained from Cone-beam computed tomography (CBCT). 
To bridge the domain gap between PX and CBCT, we create an intermediate representation called \emph{simulated PX} (SimPX). SimPX images are obtained from CBCT by simulating the process of PX imaging (Sec. Simulated PX). After creating the SimPX dataset from CBCT, we train a model for \emph{unpaired} domain translation between real PX and SimPX (Sec. Translation Module). A 3D reconstruction model, consisting of generation (Sec. Generation Module) module, is trained for generating 3D structures from 2D SimPX. At inference, real PX data is input to the translation module which generates a SimPX-style image. The image is passed to generation module to generate realistic 3D oral structures. During training, NeBLa does not require matched PX-CBCT datasets, i.e., PX and CBCT image pairs from the same patient, nor any prior information, such as the shape of patients' dental arches.

\begin{figure*}
\centering
   \includegraphics[width=1.0\linewidth]{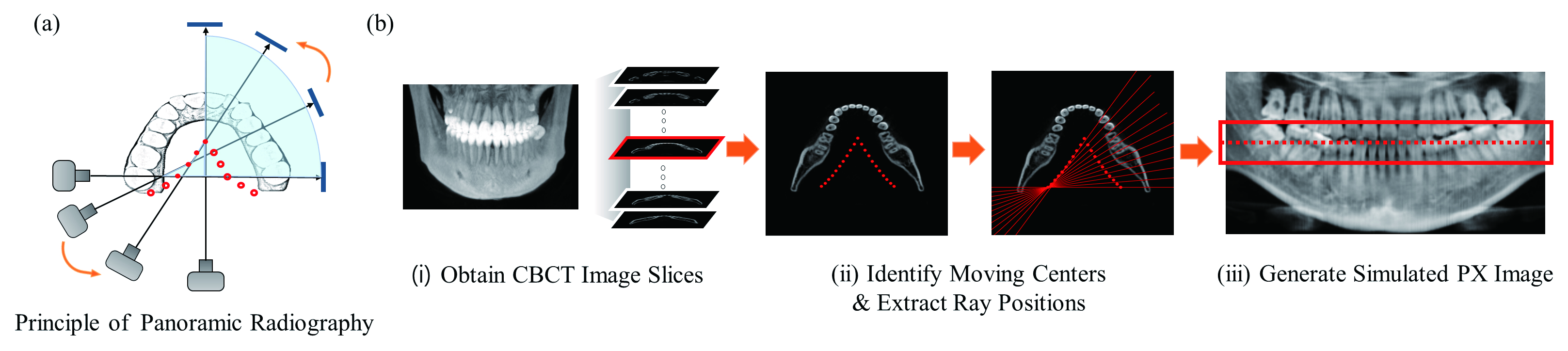}
   \caption{(a) Illustration of the process for panoramic radiography. The receptor and X-ray source rotate around the patient. The center of rotation, marked in red, moves along a curve to form a focal trough on the teeth region. (b) Process of generating SimPX. SimPX images are rendered by hypothetical rays traversing through CBCT data, whose trajectories are similar to those in (a).}
\label{fig:px}
\end{figure*}

\subsection{Background} \label{background}
PX is a diagnostic modality which provides a panoramic view of the teeth and maxillofacial skeleton in 2D images \cite{brooks2009cbct}. The principle underlying PX involves positioning the patient's head between the X-ray tube and the image receptor, capturing images as both rotate simultaneously. In order to align the focal plane with the dental curve, the center of rotation, marked as red dots in Fig. \ref{fig:px}(a), shifts as the tube and receptor rotate. This enables the formation of focal plane aligned with dental arches \cite{whaites2013essentials} which is also called \emph{focal trough}. As a result, PX shows the teeth region in sharp focus, while blurring the overlying anatomy. In this work, we take this property into account to generate simulated PX images which closely resemble real-world PX.

Cone-beam computed tomography (CBCT) is a 3D imaging technique which enables comprehensive diagnosis and evaluation of dental and maxillofacial structures. In CBCT, a cone-shaped beam of X-rays rotates around the patient capturing images at multiple angles. The projected images are processed to reconstruct 3D volumetric data, e.g., using FDK algorithm \cite{feldkamp1984practical}. CBCT is a complementary diagnostic tool to traditional PX in cases of complex dental procedures or implant placement. The downside compared to PX is the higher cost \cite{lopes2016comparison} and the exposure to higher doses of radiation \cite{brooks2009cbct}, although the dose is $10\times$ lower than conventional CT scans \cite{chen2017low}.

\subsection{Simulated PX (SimPX)} \label{simpx}
To bridge the domain gap between 2D PX and 3D CBCT, we propose an intermediate representation called SimPX (Simulated PX). The key considerations are, SimPX should be (i) ``close'' to real PX; (ii) derived directly from CBCT; (iii) ``revertible'' back to CBCT. To that end, we propose a ray tracing-based rendering of SimPX from CBCT. We hypothesize that a set of X-ray beams is cast through CBCT slices where beam directions resemble those in PX as shown in Fig. \ref{fig:px}(b). The SimPX is rendered through the estimated intensity of virtual rays traversed through CBCT.

Next, we discuss how to render SimPX images. Consider a hypothetical X-ray beam passing through a certain material. According to the Beer-Lambert law \cite{ketcham2014beam,max1995optical}, the received intensity $I$ of the X-ray, and its discrete approximation, are given by
\begin{equation}\label{eq:integral}
I = I_{0}\textrm{ exp }(-\int_{0}^{l}\mu(t)dt) \approx I_{0}\textrm{ exp }(-\sum_{i=1}^{N}\mu_{i}\epsilon_{i})
\end{equation}
where $I_{0}$ is the initial intensity, $l$ is the distance the ray traversed through the medium, $\mu$ is the linear attenuation coefficient.
In the discrete expression, $\epsilon_i$ is the infinitesimal length, and $N$ is the number of discrete samples along the ray.

In conventional CT scans, the measurements are in Hounsfield Unit (HU). HU is proportional to the linear attenuation coefficient $\mu$ given by the following relation: \cite{denotter2019hounsfield}.
\begin{equation}\label{eq:hu}
\textrm{HU} = 1000 \times \frac{\mu - \mu_{\textrm{water}}}{\mu_{\textrm{water}}}.
\end{equation}
CBCT produces lower-resolution scans with less radiation dose compared to conventional CT scans. Although CBCT images are in 4096 gray levels in the range of -1000 to +3000 similar to HU, CBCT gray values are not necessarily identical to HU. This is due to several factors such as the relatively high noise levels, the cone beam geometry, and the limited field of view (FOV) size \cite{pauwels2013variability}. Nevertheless, there exists an approximately linear relation between CBCT gray values and HU \cite{razi2014relationship, mah2010deriving}. Let $\sigma$ denote the CBCT gray value. Considering the linear relations among $\mu$, HU, and CBCT gray values, we have that $\mu \approx a\sigma+b$, where $a$ and $b$ are constants. Then the received X-ray intensity can be represented as
\begin{equation}\label{eq:intensity}
I \approx I_{0}\textrm{ exp }(-\sum_{i=1}^{N}(a\sigma_{i}+b)\epsilon_{i}) = AI_{0}\textrm{ exp }(-a\sum_{i=1}^{N}\sigma_{i}\epsilon_{i})
\end{equation}
where $A :=\textrm{ exp }(-b\sum_{i=1}^{N}\epsilon_{i}) = \textrm{ exp }(-b l)$, because the distance from the beam source to the receptor is fixed to $l$. Since $AI_0$ is constant, we can assume $AI_0=1$ after normalizing the PX images.

Next, let $\delta$ represent (infinitesimal) unit distance in the CBCT space. We may assume that the physical distance is proportional to the CBCT distance, i.e., $\delta_i = c\epsilon_i$ for some constant $c$. Thus, from Eq. \ref{eq:intensity}, the transmittance \cite{max1995optical} of the ray is given by
\begin{equation}\label{eq:trans}
T=\exp\bigg[-\sum_{i=1}^{N}(\beta\sigma_{i})\delta_{i}\bigg]
\end{equation}
where $\beta:=a/c$. Eq. \ref{eq:trans} is equivalent to the volume rendering with absorption only. The pixel value of X-ray corresponds to $1-T$ or \emph{opacity} \cite{max1995optical}. 

The SimPX is generated from CBCT according to Eq. \ref{eq:trans}. In the CBCT domain, the attenuation coefficient is effectively the CBCT gray values scaled by $\beta$ from Eq. \ref{eq:trans}. Since $\beta$ is unknown, we set $\beta$ using a hyperparameter search.
The virtual X-ray beams are arranged to traverse the CBCT where each beam gives the rendered value of $1-T$ in the SimPX. The trajectories of rays are chosen to generate realistic PX images as depicted in Fig. \ref{fig:px} (b). For example, the rays passing near molars are sampled more than those closer to incisors. The detailed configuration of generating SimPX is given in Appendix.% \ref{simpx_detail}.

\subsection{Translation Module} \label{translation}
Although SimPX is generated based on the principles of X-ray imaging, there still exists a domain gap between real PX and SimPX.
For example, the exact trajectory of the center of rotation of the PX tube is typically unknown and varies with the manufacturer \cite{whaites2013essentials}. In SimPX, we heuristically use a quadratic curve for the center trajectory: see Appendix.% \ref{simpx_detail}.
In addition, real PX images may have artifacts, e.g., due to overlying anatomy such as cervical vertebrae or the movement of patients \cite{whaites2013essentials}. 

To bridge this modality gap, we consider a domain translation between real PX and SimPX images. Importantly, we perform \emph{unpaired} image-to-image translation, and use CycleGAN \cite{zhu2017unpaired} as the base model.  Firstly, SimPX images are generated from the CBCT dataset. Next, we train the model for unpaired translation between SimPX and real PX datasets. Thus, the CBCT data need not be from the same patients as PX data. This approach has practical significance because the matched PX and CBCT data is hard to obtain in clinical practice. 

We consider regularization based on semantic information for improved translation. In PX images, the teeth region is shown in higher contrast than the overlying anatomy. Thus, we propose to guide translation with the semantic information on teeth, so that the teeth shapes are well-preserved during translation. To achieve this, we consider regularization with \emph{semantic consistency} as follows.
A teeth segmentation model is trained for PX images, i.e., the model outputs a binary prediction of whether a pixel belongs to the teeth region. A UNet-based \cite{unet} segmentation model $S$ is trained using a publicly available  PX dataset \cite{abdi2015automatic}. Given a real PX image $x$, we want $x$ and the image translated from $x$ produce similar segmentation results. The loss functions are given by
\begin{equation}
\begin{gathered}
     \mathcal{L}= \mathcal{L}_{\textrm{CycleGAN}}(G, H, D_X, D_{Y}) + \lambda \mathcal{L}_{\textrm{seg}}(G), \\
     \quad\mathcal{L}_{\textrm{seg}}(G)= \mathbb{E}_{x \sim p_{X}(\cdot)}[\ell(S(G(x)),S(x))]
\end{gathered}
\end{equation}
where $G$ (resp.\ $H$) is the translation model from real PX to SimPX (resp.\ from SimPX to real PX); $X$ (resp.\ $Y$) is the domain of real PX (resp.\ SimPX); $D_X$ and\ $D_Y$ are discriminators; $\ell$ is the pixel-wise MSE loss; $S(x)$ and $S(G(x))$ denote the sigmoid output values for segmentation of real PX image $x$ and translated image $G(x)$ respectively. The semantic consistency regularization gives a clearer generation of teeth structure, and the experimental study is in Appendix.% \ref{translation_ablation}.

\begin{figure*}[t!]
\centering
   \includegraphics[width=1.0\linewidth]{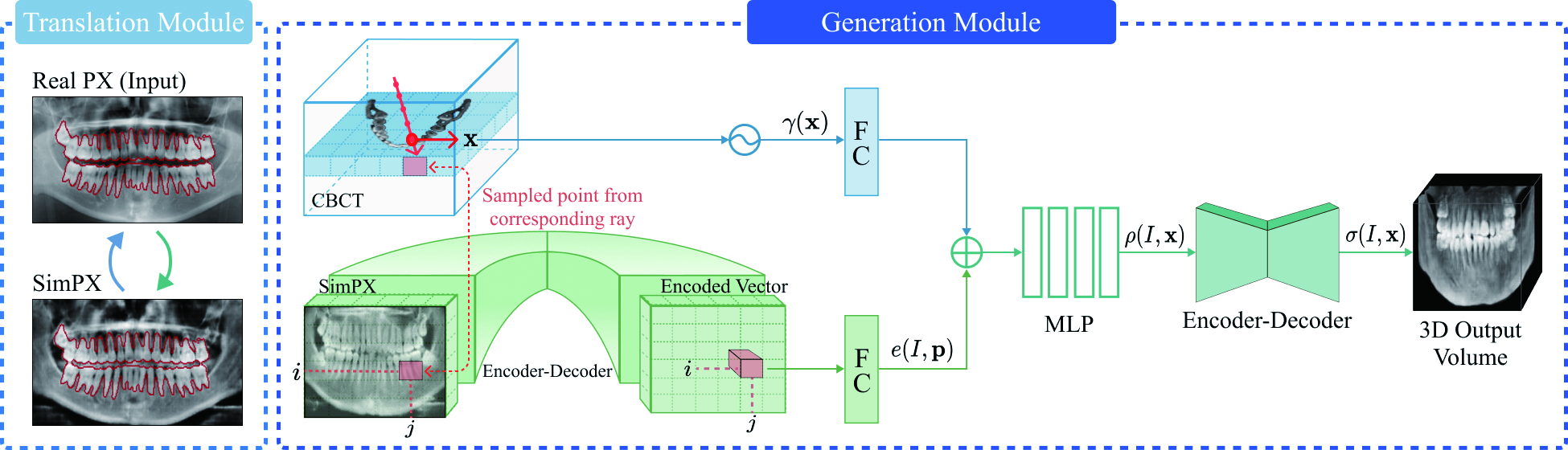}
   \caption{Overview of NeBLa. Function $\gamma(\mathbf{x})$ denotes the positional encoding applied to position $\mathbf{x}$ sampled from a ray, as in \cite{mildenhall2020nerf}. FC stands for the fully connected layer.}
\label{fig:model}
\end{figure*}

\subsection{Generation Module} \label{generation}
We introduce generation module which reconstructs CBCT values from SimPX. From Eq.\ \ref{eq:trans}, gray value $\sigma$ can be regarded as (scaled) material density, % for rendering.
and thus will be called the density in this section. We consider a neural model for density prediction inspired by NeRF \cite{mildenhall2020nerf}; however, our model takes input images from \emph{multiple} objects (patients) for direct prediction of density unlike NeRF or its variants for view synthesis of a \emph{single} object. 
Our model is a function $F$ defined as
\begin{equation}\label{eq:F}
F(\mathbf{x}, \mathbf{e}(I, \mathbf{p})) 
\end{equation}
where $F$ returns a density estimate at voxel location $\mathbf{x}$ conditional on the ray parameterized by $\mathbf{p}$, input SimPX image $I$ and the associated embedding $\mathbf{e}(I,\mathbf{p})$. We use an MLP model for $F$. The details of the model are explained as follows. 

\textbf{Point Sampling from Rays.}  $I$ is the SimPX image rendered from CBCT based on Eq. \ref{eq:trans}. $\mathbf{p} = (i,j)$ represents the position of pixel at $i$-th row and $j$-th column of $I$. The ray used to render the pixel at $\mathbf{p}$ has a predetermined starting position $\mathbf{o}$ and direction $\mathbf{d}$ which are given by the rotating ray trajectories in SimPX. Thus, the path of a ray is a line $\mathbf{o}(\mathbf{p}) + t\cdot\mathbf{d}(\mathbf{p})$ parameterized by $t$ where
$\mathbf{x}$ is a sampled position on the ray. The sampling interval is given by $\delta_i$ which also is the distance parameter of the transmittance in Eq.\ \ref{eq:trans}. The interval is set uniformly in our model.

\textbf{Image and Position Embedding.} The density at $\mathbf{x}$ contributes to position $\mathbf{p}$ of rendered image $I$ through the ray passing $\mathbf{x}$.
To account for this, we use image and position embedding $\mathbf{e}(I, \mathbf{p})$ as input to $F$.
An encoder-decoder model is used to obtain embedding $\mathbf{e}(I, \mathbf{p})$, specifically UNet \cite{unet}.
The spatial dimension of input and output can be made identical in UNet, preserving the spatial locality of input image features. We set $\mathbf{e}(I,\mathbf{p})$ as the feature vector at position $\mathbf{p}$ at the UNet output, which corresponds to the encoded features of the same position $\mathbf{p}$ at input image $I$ as illustrated in Fig. \ref{fig:model}. Moreover, the encoder-decoder architecture of UNet enables capturing both the local and global features of input images. This is crucial for our model, because $F$ is trained over different objects (patients), and needs to learn the density distribution over input images.

\textbf{Density Estimation.} The rays used for generating SimPX are nonparallel, and multiple rays may intersect at the same voxel. Thus, it is somewhat ambiguous to regard $F$ as the eventual density at $\mathbf{x}$, because there may exist another ray crossing $\mathbf{x}$ providing a different density estimate at $\mathbf{x}$. Therefore, we define the \emph{intermediate} density at $\mathbf{x}$ given input image $I$, denoted by $\rho(I, \mathbf{x})$, as
\begin{equation}\label{eq:ino}
\rho(I,\mathbf{x}) =
\frac{1}{|B(\mathbf{x})|} \sum_{\mathbf{p} \in B(\mathbf{x})} F(\mathbf{x}, \mathbf{e}(I,\mathbf{p}))
\end{equation}
where $B(\mathbf{x})$ denotes the set of pixel positions $\mathbf{p}$ such that rays incident on $\mathbf{p}$ crosses the voxel at $\mathbf{x}$. The intermediate density $\rho$ is a crude estimate based on the average of multiple (candidate) density values. $\rho$ will be processed by another function for the eventual density, where the function is learned from the distribution of 3D structures over different cases (patients). Another view on Eq. \ref{eq:ino} is as follows. The ray trajectories of SimPX result in spatially \emph{nonuniform sampling} of the density field. Since $|B(\mathbf{x})|$ quantifies how frequently $\mathbf{x}$ is sampled by rays, sparsely sampled locations are given higher weights in learning $F$ by Eq. \ref{eq:ino}, analogous to inverse-frequency class weighting \cite{huang2016learning}.

The intermediate density $\rho(I,\mathbf{x})$ in Eq. \mbox{\ref{eq:ino}} is input to an encoder-decoder function to yield the final density $\sigma(I,\mathbf{x})$. The encoder-decoder layer (a U-Net) refines the crude estimate $\rho$ by learning 3D structures from various cases of multiple patients.

\textbf{Loss functions.} We apply three types of loss function to the output: (i) MSE loss $\mathcal{L_{\textrm{MSE}}}$; (ii) projection loss $\mathcal{L}_{\textrm{proj}}$ characterized by the MSE computed across the maximum intensity projection (MIP) images oriented along the axial, sagittal, and coronal planes; (iii) perceptual loss $\mathcal{L}_{\textrm{perc}}$ \mbox{\cite{johnson2016perceptual}} between the predictive output and the ground truth.

The loss is applied to the density prediction $\sigma(I,\mathbf{x})$ relative to the ground truth CBCT values. Specifically,
\begin{equation}\label{eq:nebla_loss}
\mathcal{L} = \mathcal{L}_{\textrm{MSE}} + \lambda_1 \mathcal{L}_{\textrm{proj}} + \lambda_2 \mathcal{L}_{\textrm{perc}},\qquad
\end{equation}
\begin{equation}\label{eq:mse}
\mathcal{L_{\textrm{MSE}}} = \mathbb{E}_{I\sim p_I(\cdot)}\bigg[ \sum_{\mathbf{x}\in\mathbb{R}^3} (\sigma(I,\mathbf{x})-\hat\sigma(I,\mathbf{x}))^2\bigg]
\end{equation}
where $\hat{\sigma}$ represents the ground truth CBCT values which are used to generate SimPX image $I$. The expectation is taken over SimPX images generated from the CBCT dataset. $\mathcal{L}_{\textrm{proj}}$ and $\mathcal{L}_{\textrm{perc}}$ can be defined similar to  $\mathcal{L_{\textrm{MSE}}}$.

\textbf{Differences from NeRF.}
Since our objective is to generate a 3D output based on the PX image, the images from different objects (patients) are input to the model at each training step.
In NeRF \cite{mildenhall2020nerf}, the objective is the synthesis of novel views to a single object. The NeRF model predicts the color and density given viewing direction, and the loss function is given by
\begin{equation}\label{eq:nerf_loss}
\mathcal{L} = \mathbb{E}_{\Pi\sim p_\Pi(\cdot)}\bigg[ \sum_{r\in R(\Pi)} \|\mathbf{c}(r)-\hat{\mathbf{c}}(r)\|^2\bigg]
\end{equation}
where $\mathbf{c}$ (resp. $\hat{\mathbf{c}}$) denotes the rendered (resp. ground truth) pixel value, and $R(\Pi)$ denotes the set of rays associated with camera pose $\Pi$. Unlike Eq. \mbox{\ref{eq:mse}}, the expectation is taken over a distribution of camera poses around the fixed object. Some variants of NeRF including pixelNeRF \cite{yu2021pixelnerf} also propose the density estimation conditional on the rendered images, but the models use similar loss functions so as to generate novel views of a single object.

\begin{table*}[t]
\begin{center}
\setlength\tabcolsep{5pt}
\fontsize{9pt}{9pt}\selectfont
\def\arraystretch{1.3}%
\begin{tabular}{|l|c|c|c|c|c|c|c|c|}
\hline
\multicolumn{1}{|c|}{} &  \multicolumn{4}{c|}{Real PX$\Rightarrow$CBCT} & \multicolumn{4}{c|}{Synth.\ PX$\Rightarrow$CBCT}\\
\hline
Method & PSNR (dB) & SSIM (\%) & Dice (\%) & LPIPS(VGG) & PSNR (dB) & SSIM (\%) & Dice (\%) & LPIPS(VGG)\\
\hline
Residual CNN & 17.91$\pm$0.12 & 65.22$\pm$1.70 & 44.85$\pm$7.19 & 0.49$\pm$0.006 & 18.40$\pm$0.11 & 68.20$\pm$1.23 & 57.16$\pm$1.18 & 0.46$\pm$0.009\\
GAN & 18.02$\pm$0.12 & 66.00$\pm$2.10 & 46.84$\pm$6.79 & 0.50$\pm$0.010 & 18.55$\pm$0.09 & 68.20$\pm$1.49 & 57.18$\pm$1.34 & 0.48$\pm$0.009\\
R2N2 & 17.23$\pm$0.38 & 77.32$\pm$1.16 & 56.54$\pm$4.48 & 0.34$\pm$0.004 & 17.76$\pm$0.26 & 78.22$\pm$0.60 & 53.86$\pm$4.05 & 0.33$\pm$0.002\\
X2CT-GAN & 18.55$\pm$0.24 & 77.43$\pm$1.85 & 58.85$\pm$6.36 & 0.34$\pm$0.030 & 18.41$\pm$0.40 & 77.05$\pm$2.20 & 62.55$\pm$1.24 & 0.32$\pm$0.030\\
Oral-3D & 17.11$\pm$0.09 & 72.96$\pm$0.55 & 50.68$\pm$1.07 & 0.44$\pm$0.004 & 17.66$\pm$0.06 & 74.83$\pm$0.12 & 52.08$\pm$0.20 & 0.42$\pm$0.002\\
\textbf{NeBLa (Ours)} & \textbf{19.89}$\pm$\textbf{0.30} & \textbf{79.21}$\pm$\textbf{0.64} & \textbf{65.22}$\pm$\textbf{2.72} & \textbf{0.30}$\pm$\textbf{0.007} & \textbf{21.68}$\pm$\textbf{0.25} & \textbf{82.62}$\pm$\textbf{0.34} & \textbf{74.77}$\pm$\textbf{1.11} & \textbf{0.24}$\pm$\textbf{0.006}\\
\hline
\end{tabular}
\end{center}
\caption{Quantitative comparison of 3D reconstruction using real-world and synthesized PX. The format is \emph{mean}$\pm$\emph{std} with 10 repetitions of experiments.}
\label{tab1}
\end{table*}

\begin{figure*}[h]
\centering
\includegraphics[width=1.0\linewidth]{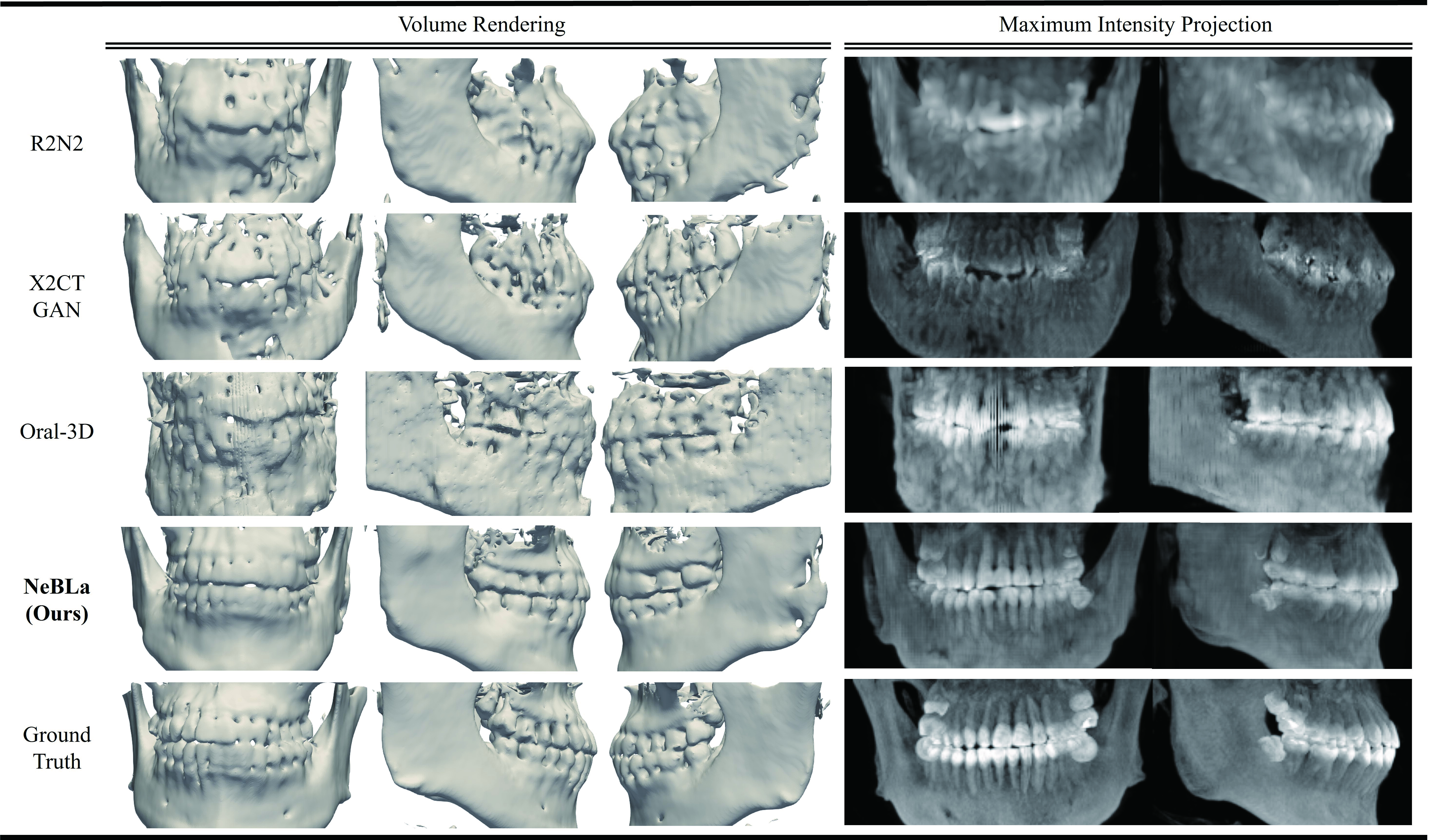}
   \caption{Qualitative comparison of 3D reconstruction results from real PX image using volume rendering and maximum intensity projection.}
\label{fig:vis}
\end{figure*}

\begin{figure}[h]
\centering
\includegraphics[width=1.0\linewidth]{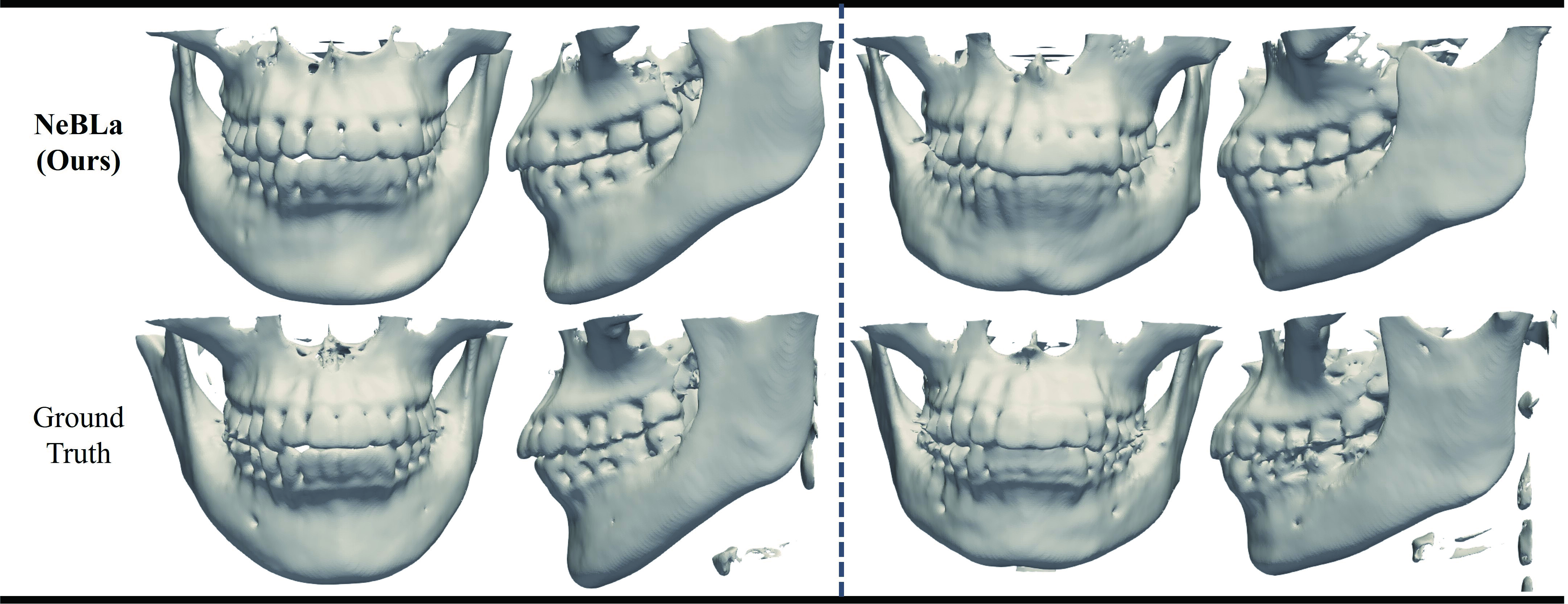}
   \caption{Qualitative comparison of 3D reconstruction results from SimPX images of two different patients using volume rendering.}
\label{fig:vis2}
\end{figure}

\section{Experiments}
\subsection{Dataset}
For training the translation module, SimPX dataset is generated from 44 cases of CBCT data, and unpaired real PX dataset of 45 cases is used. For training the remaining module of NeBLa and the baseline models, 90 cases of CBCT are used with 55 cases for training and 4 cases for validation. 31 cases of paired CBCT and real-world PX are used for testing. All the datasets were obtained from Korea University Anam Hospital. This study received approval from the Institutional Review Board at Korea University (IRB number: 2020AN0410).

\subsection{Baselines and Evaluation Metrics}
We compare NeBLa with baselines including the state-of-the-art (SOTA) methods which address similar task. Oral-3D \cite{song2021oral}  utilized the GAN model to generate 3D oral structures from a single PX image. R2N2 \cite{choy20163dr2n2} used a Recurrent Neural Network (RNN) to reconstruct 3D shapes from multi-view images. In our case, the recurrent connection was removed because the number of input images is one. Residual CNN \cite{henzler2018single} used an encoder-decoder network for reconstruction. For baseline GAN  \cite{goodfellow2020generative}, we used Residual CNN as the generator and the same discriminator as Oral-3D. X2CT-GAN \cite{ying2019x2ct} used GAN to reconstruct 3D CT volumes from 2D biplanar X-rays where we used the single-view setting of their model. The training settings for models are provided in Appendix.% \ref{implementation}.

The metrics for assessing the quality of 3D outputs are: Peak Signal-to-Noise Ratio (PSNR), Structural Similarity Index (SSIM), and Dice coefficient. We also use Learned Perceptual Image Patch Similarity (LPIPS) comparing generated and target images based on perceptual similarity \cite{zhang2018unreasonable}. For LPIPS, we use feature outputs from pre-trained models, e.g., AlexNet, VGG-16, and SqueezeNet.

\subsection{Results}
While our main task is to predict CBCT from real PX, some challenges exist in the performance evaluation. A pair of PX and CBCT taken from the same patient are used as input and target. However, PX and CBCT are taken at different times, and patient's pose or dental conditions may differ between modalities \cite{white2014oral}. This fundamentally limits the exact prediction of CBCT from the paired PX. We consider two modes of evaluation to account for the limitation. 

\textbf{Real PX$\Rightarrow$CBCT.} This mode is relevant to the main task of this paper: reconstructing CBCT from real PX. Real PX is translated to some synthesized-style PX image (depending on the model), from which 3D structure is reconstructed. The target CBCT is not exact, but approximate, ground truth.

\begin{table}
\begin{center}
\fontsize{9pt}{9pt}\selectfont
\def\arraystretch{1.3}%
\begin{tabular}{l|ccc}
    \hline
    LPIPS & AlexNet & VGGNet & SqueezeNet\\
    \hline
    \textbf{Ours} & \textbf{0.24}$\pm$\textbf{0.02} & \textbf{0.31}$\pm$\textbf{0.02} & \textbf{0.18}$\pm$\textbf{0.01}\\
    SynA & 0.40$\pm$0.02 & 0.43$\pm$0.01 & 0.32$\pm$0.01\\
    SynB & 0.33$\pm$0.02 & 0.41$\pm$0.02 & 0.27$\pm$0.01\\
    \hline
\end{tabular}
\end{center}
\caption{Comparison of perceptual similarity between real and translated PX images. A lower value indicates higher perceptual similarity.}
\label{tab2}
\end{table}

\begin{figure}[h]
\centering
\includegraphics[width=1.0\linewidth]{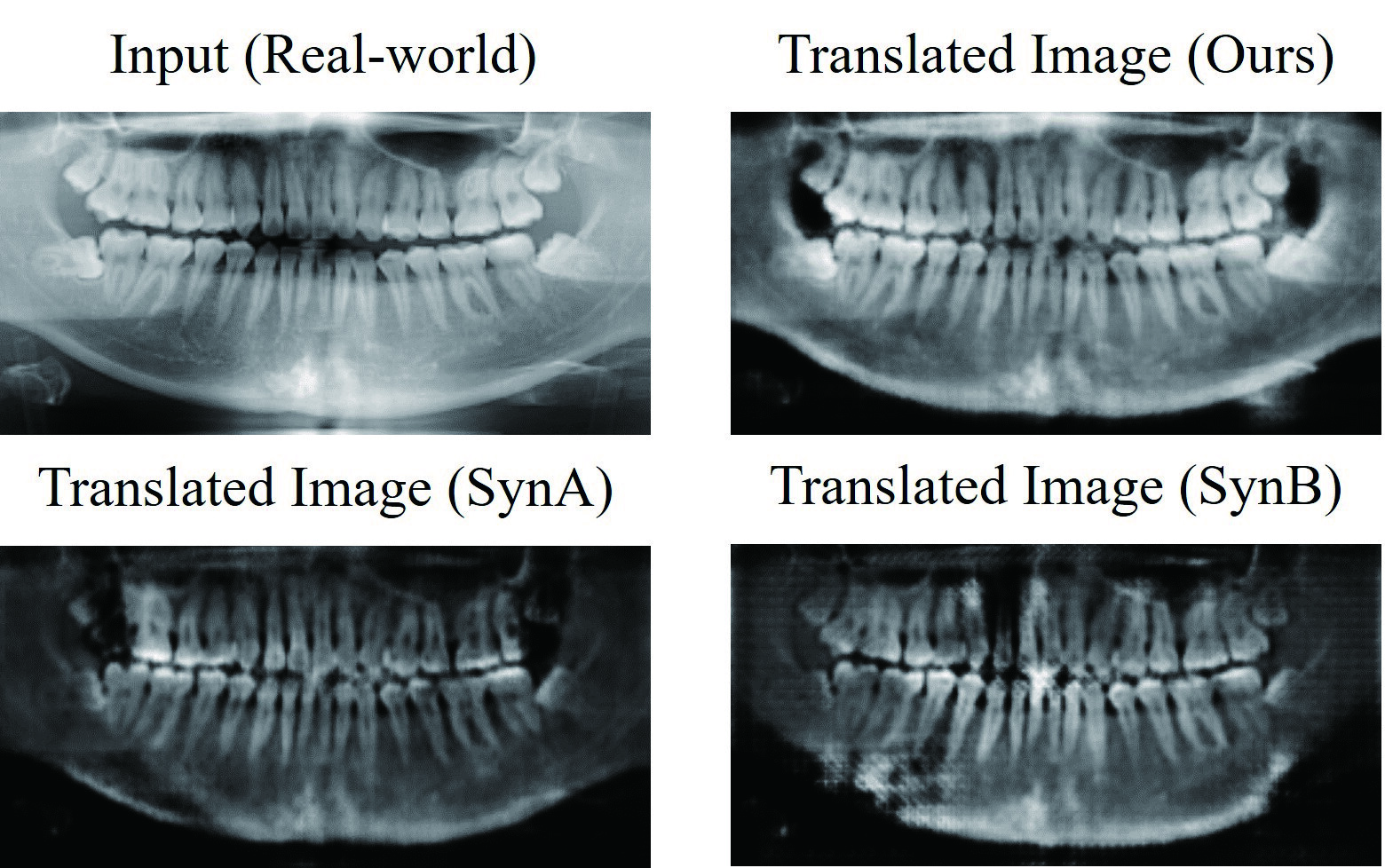}
   \caption{Qualitative comparison of translated PX images with real PX.}
\label{fig:cyclegan}
\end{figure}

\textbf{Synth.\ PX$\Rightarrow$CBCT.} In this mode, we synthesize a PX image from a given CBCT, where the synthesis method may differ depending on the model. The image is reverted back to CBCT. This mode better evaluates the reconstruction capability of the model, because ground truth (CBCT) exists as the target. %form 2D of our ray-based approach 

We first consider evaluation mode \emph{Real PX$\Rightarrow$CBCT}. The quantitative results are shown in the left columns of Table \ref{tab1}, and the qualitative results are in Fig. \ref{fig:vis}. Table \ref{tab1} shows that NeBLa significantly outperforms the baselines. Fig. \ref{fig:vis} shows that NeBLa produces results which are substantially closer to the CBCT data. In particular, NeBLa is significantly more accurate in generating the teeth part, which is crucial for dental applications. Although visibly much superior to the baselines, the intrinsic misalignment between PX and CBCT seems to limit the performance of NeBLa. One may expect larger gaps in quantitative results, in the settings with better aligned modalities as follows.

Next, we consider evaluation mode \emph{Synth.\ PX$\Rightarrow$CBCT} which provides an alternative view on the reconstruction performance. For synthesizing PX images, NeBLa uses SimPX, and the baseline models use the method proposed by \cite{yun2019automatic} which extracts dental arch from maximum intensity projection, and synthesizes the curved images along the arch. As shown in the right columns of Table \ref{tab1}, the performance gap between NeBLa and baselines is significantly widened compared to mode \emph{Real PX$\Rightarrow$CBCT}. This shows that NeBLa is able to produce even more accurate outputs when the exact ground truth is available, manifesting its reconstruction capabilities. Fig. \ref{fig:vis2} shows the qualitative results, which support the quantitative results. More visualization results are provided in Appendix.% \ref{more_visual}.

It is important that the synthesized PX images have a similar style to real PX to facilitate domain translation. We evaluate the similarity of the proposed SimPX to real PX images. SimPX is compared with two existing methods \cite{yun2019automatic}, \cite{amorim2020reconstruction} for synthesizing PX from CBCT which is denoted by SynA and SynB respectively. SynA is also used for synthesizing PX for baseline models in previous experiments. SynB fits B{\'e}zier curves to the segmented dental arch and synthesizes the PX image with parallel B{\'e}zier curves.
Three translation modules are trained, one for each synthesis method, and the real PX and translated PX images from each module are compared.

Table \ref{tab2} shows that SimPX produces translated images closer in style to the real PX, indicating a relatively small gap between real PX and SimPX domains. Fig. \ref{fig:cyclegan} shows a qualitative comparison of synthesis methods. The SimPX yields the most accurate and visually plausible results, whereas other synthesizing methods show considerable distortion in the teeth region. We hypothesize that this is due to the fact that SynA and SynB extract the dental arch from CBCT, and generate synthesized images using curved projections, which are not directly related to the PX imaging method. In contrast, SimPX incorporates the principles of PX imaging into the synthesis process, resulting in a smaller gap between real PX and SimPX domains.

\textbf{Ablation Study.} Ablation study of NeBLa is provided in Appendix.

\section{Conclusions}
In this paper, we introduce a novel architecture for reconstructing 3D oral structures from real-world PX images. To our knowledge, NeBLa is the first attempt to generate 3D oral structures from real-world PX without any prior information. Importantly, our framework is capable of producing high-quality oral structures and is more robust to artifacts in real-world PX than other state-of-the-art methods. This capability makes NeBLa valuable for supporting educational tools based on AR/VR or supplementary materials for 3D visualization. A limitation of our work is the effect of the unknown trajectory of the center of rotations of X-ray beams in PX images. Although we resorted to heuristics and domain translation to mitigate the effect, we plan to devise algorithms to better estimate the trajectories adapting to the variations in equipment.

\section{Acknowledgements}
This work was partly supported by the Korea Medical Device Development Fund grant funded by the Korea Government (the Ministry of Science and ICT, the Ministry of Trade, Industry and Energy, the Ministry of Health \& Welfare, the Ministry of Food and Drug Safety) (Project Number: 1711195279, RS-2021-KD000009, 25\%); the National Research Foundation of Korea (NRF) Grant through the Ministry of Science and ICT (MSIT), Korea Government, under Grant 2022R1A5A1027646 (25\%); the National Research Foundation of Korea (NRF) grant funded by the Korea Government(MSIT) (No. 2021R1A2C1007215, 25\%); ICT Creative Consilience Program through the Institute of Information \& Communications Technology Planning \& Evaluation(IITP) grant funded by the Korea government(MSIT) (IITP-2024-2020-0-01819, 25\%)

\bibliography{aaai24}
\clearpage
\normalsize

\renewcommand\thesubsection{\Alph{subsection}}

\section*{Appendix}

\subsection{Procedures to generate SimPX images} \label{simpx_detail}

\begin{figure*}[t!]
\begin{center}
\includegraphics[width=0.85\linewidth]{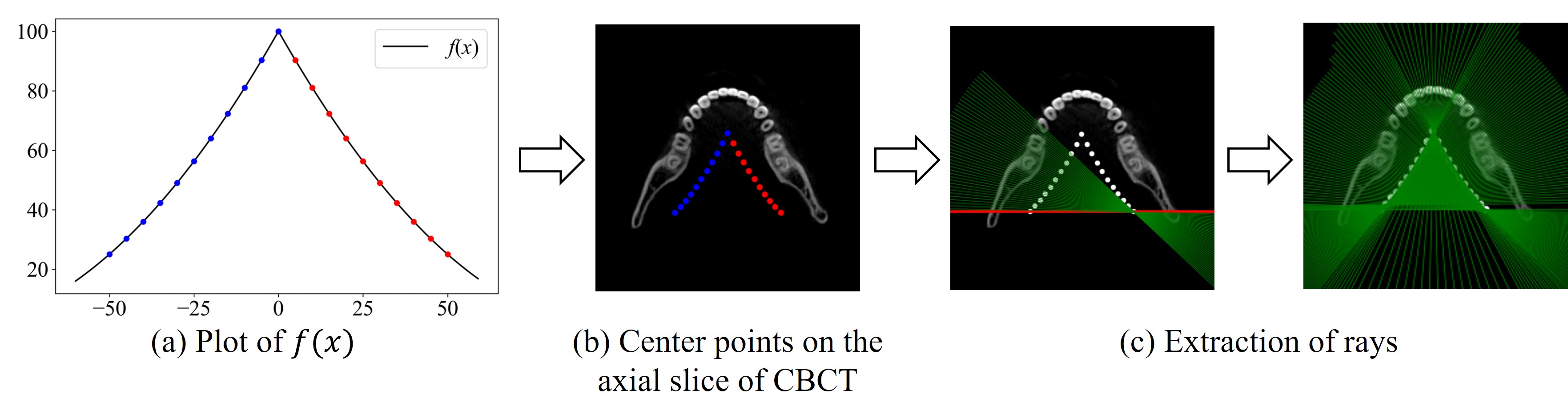}
\end{center}
\caption{Process of ray extraction}
\label{fig:appenA}
\end{figure*} 

\paragraph{Step1: Rotation Centers.}
As outlined in Sec. \ref{background}, the focal trough in PX imaging is determined by the trajectory of the center of rotation of the PX tube and the receptor. However, obtaining the precise trajectory from real PX images is difficult, because standard trajectories do not exist. Moreover, the trajectory varies with and is often proprietary information, to the equipment manufacturers. In this work, we heuristically use a quadratic curve for the center trajectory. Our experiments showed that, the trajectory given by a quadratic curve of the form \[f(x)=\left\{
\begin{array}{cc}
0.01(x+100)^2,&-100\leq x \leq 0\\
0.01(x-100)^2,&0\leq x \leq 100
\end{array}\right.\] yielded good results. A plot of $f(x)$ is given in Fig. \ref{fig:appenA}(a). The curve will generate center locations which will be placed on the axial slice of CBCT of size $[0, 256] \times [0, 256]$. Specifically, a total of 21 points on the curve are sampled, and $i$-th point denoted by $c_i$ is given by \[
c_i=\left(5i-50,f(5i-50)\right),\quad i=0,\ldots,20.\]
In other words, 21 points on curve $f(x)$ are sampled for $-50\leq x \leq 50$ with the interval on the $x$-axis of 5. $c_i$ are placed on the axial slice of CBCT, and serve as the center points of the rays, e.g., see Fig. \ref{fig:appenA}(b) and (c). The rays will be generated from $c_i$ as follows.

\begin{figure}[h]
\begin{center}
\includegraphics[width=1.0\linewidth]{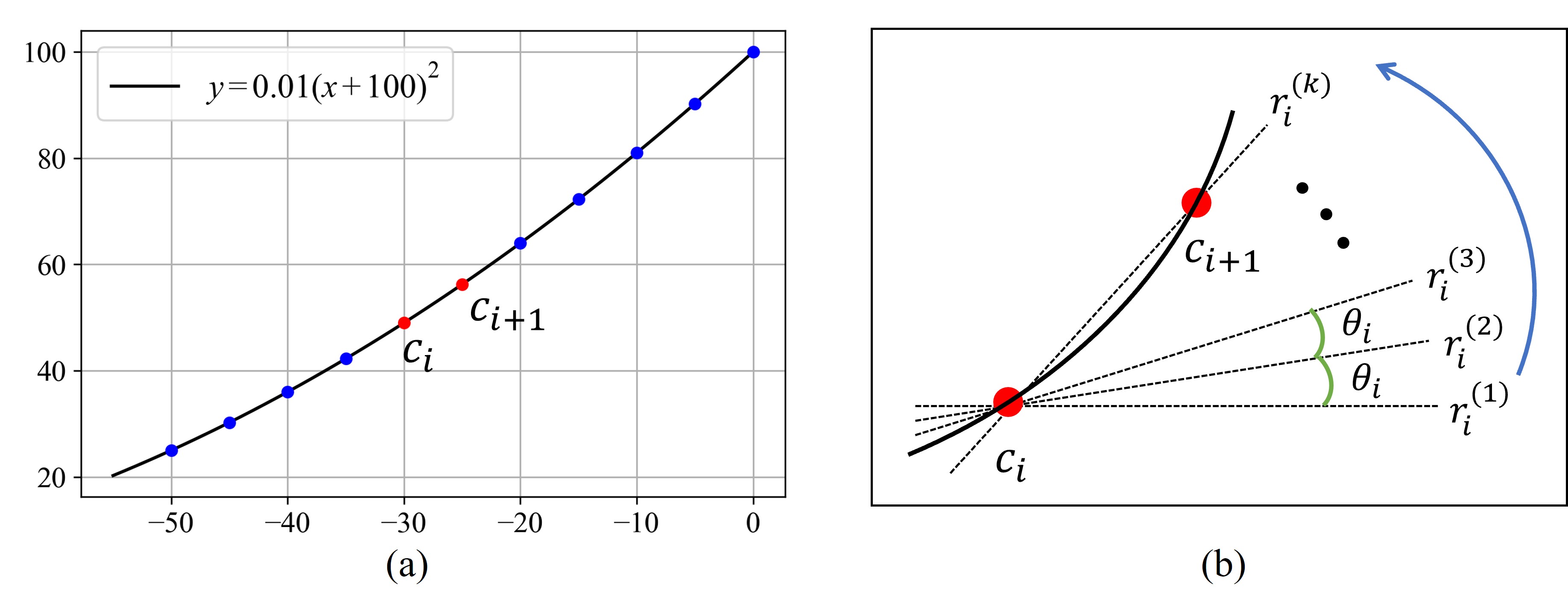}
\end{center}   
\caption{(a) Examples of center locations $c_i$ and $c_{i+1}$ on the curve $f(x)$. (b) Sampling rays  between two adjacent center points  $c_i$ and $c_{i+1}$.}
\label{fig:appenA2}
\end{figure} 

\paragraph{Step2: Ray Extraction.}
The rays are sampled by traversing center points. Consider two center points $c_i$ and $c_{i+1}$ in Fig. \ref{fig:appenA2}(a).

Starting from $c_i$, rays are sampled by rotating the ray by angle $\theta_i$, which results in rays labeled as $r_{i}^{(1)}, ..., r_{i}^{(k)}$. The final ray $r_{i}^{(k)}$ is obtained by connecting $c_i$ and $c_{i+1}$ as shown in Fig. \ref{fig:appenA2}(b). Next, we move to $c_{i+1}$ which is the new center point for rotation, and continue sampling in a similar manner. The sampling angle $\theta_i$ changes with $i$, which is given by
\begin{equation*}
\theta_i =\begin{cases}
0.5^{\circ} &\text{if $i=0, 1, 18, 19$}\\
1.5^{\circ} &\text{if $i=10$}\\
0.6^{\circ} &\text{otherwise}
\end{cases}
\end{equation*}
The sampling angles change depending on the location of center points in order to  generate realistic PX images. We adjust the number of sampled rays based on their proximity to different dental regions. For example, more rays are sampled near the molars, while fewer rays are sampled near the incisors.

\begin{figure}[h]
\begin{center}
\includegraphics[width=1.0\linewidth]{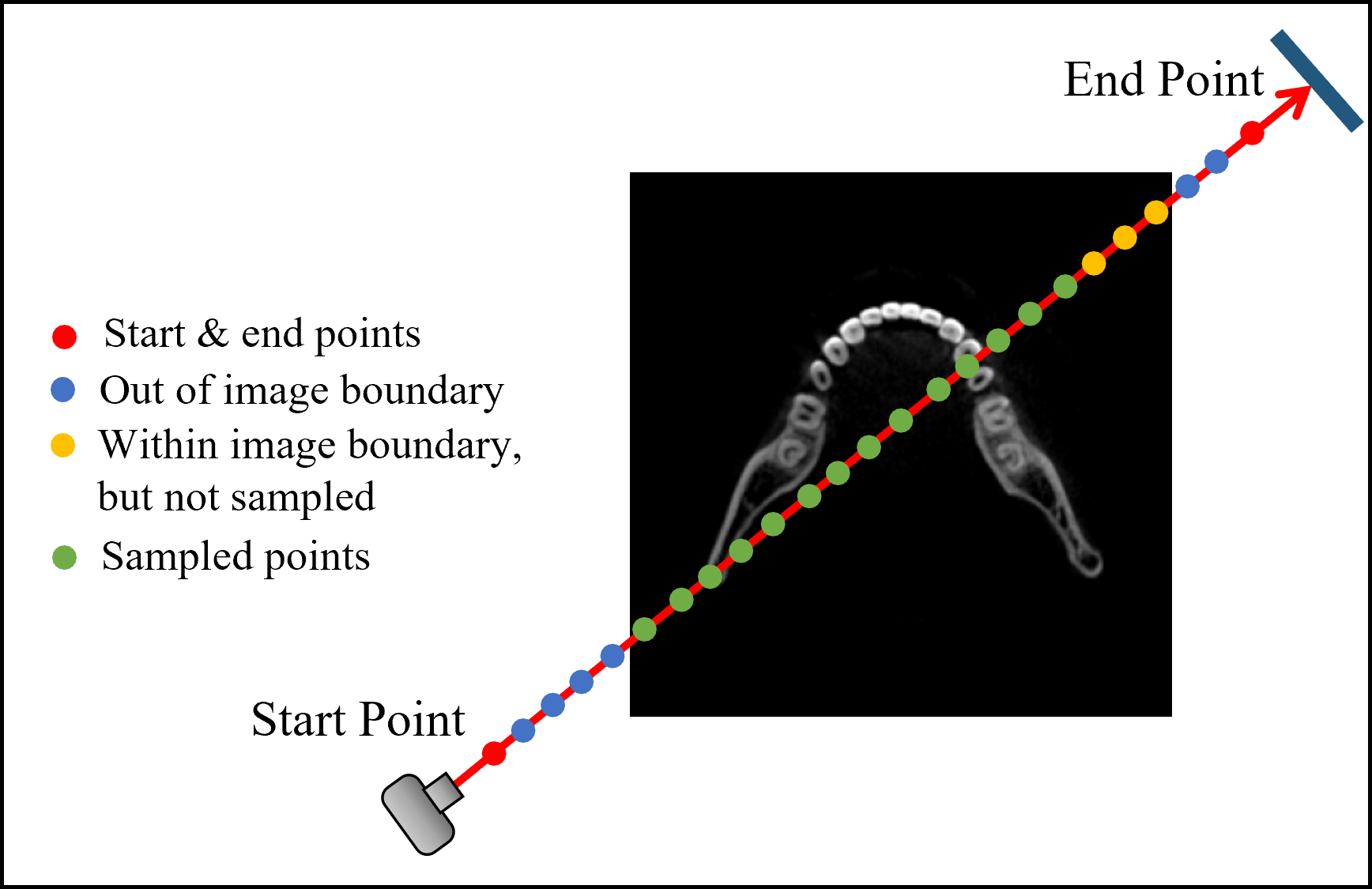}
\end{center}   
\caption{Sampled points from a ray. The interval between sampled points is uniform and identical across all rays.}
\label{fig:appenA3}
\end{figure} 

\paragraph{Step3: Point Sampling and Pixel Rendering.}
After extracting the rays, we sample points along the rays. For each ray, 200 points are uniformly sampled as follows. Consider the start and end point of a ray, as illustrated by the red points in Fig. \ref{fig:appenA3}. We proceed to sample points along the ray at equal intervals, ensuring that only points falling within the boundary of the image are selected. These sampled points are depicted as green and yellow points in Fig. \ref{fig:appenA3}. To maintain a fixed number of sampled points, we select the first 200 points counting from the start point. Consequently, only the green points shown in Fig. \ref{fig:appenA3} are retained as the final set of sampled points. These points are then rendered to a single pixel using Eq. \eqref{eq:trans}.

\subsection{Ablation Analysis} \label{ablation}
\subsubsection{Translation Module} \label{translation_ablation}

\begin{figure*}[h]
\begin{center}
\includegraphics[width=1.0\linewidth]{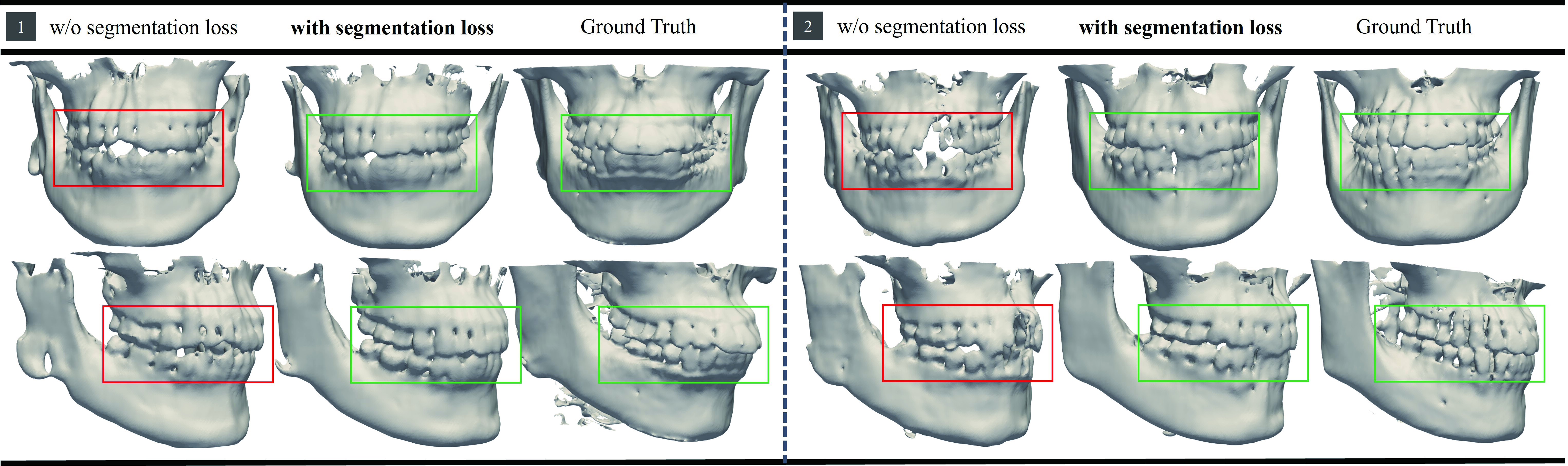}
\end{center}
\caption{Qualitative comparison of 3D reconstruction results for Case 1 and 2 with respect to the use of segmentation loss in the translation module.}
\label{fig:sup_vis_seg}
\end{figure*} 

\begin{table*}[h]
\begin{center}
\setlength\tabcolsep{9pt}
\begin{tabular}{|l|c|c|c|c|}
\hline
Method & PSNR (dB) & SSIM (\%) & Dice (\%) & LPIPS\\
\hline
w/o segmentation loss& 19.71$\pm$0.20 & 78.19$\pm$0.41 & 66.08$\pm$1.37 & \textbf{0.304}$\pm$\textbf{0.006}\\
with segmentation loss & \textbf{19.72}$\pm$\textbf{0.21} & \textbf{78.58}$\pm$\textbf{0.35} & \textbf{67.00}$\pm$\textbf{1.11} & 0.306$\pm$0.006\\
\hline
\end{tabular}
\end{center}
\caption{Ablation study on the segmentation loss in translation module.}
\label{cyclegan_ablation}
\end{table*}

\begin{figure}[h]
\centering
\includegraphics[width=1.0\linewidth]{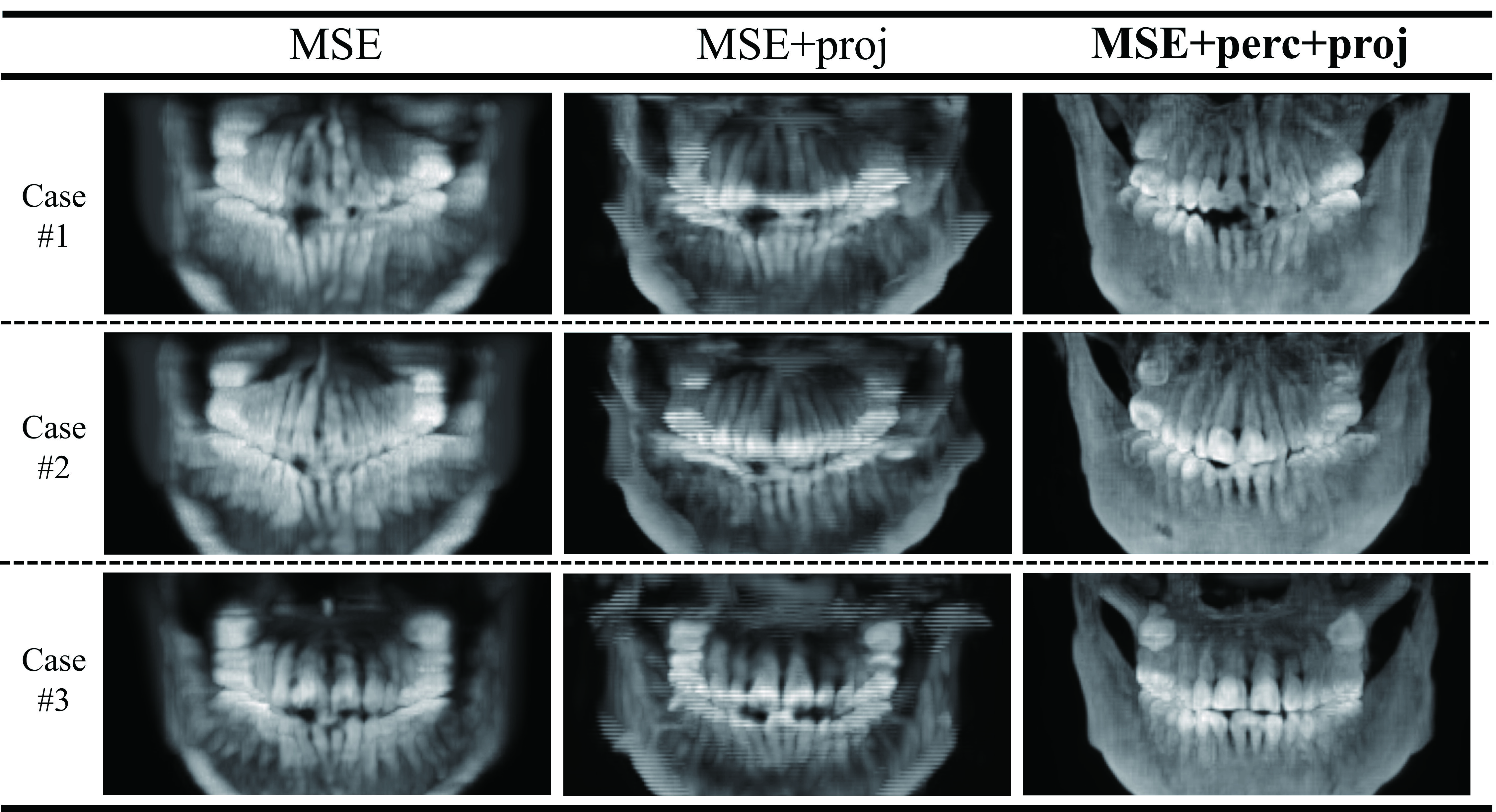}
   \caption{Qualitative comparison of 3D reconstruction results with different combinations of loss functions. ``MSE'' stands for the MSE loss only, ``MSE+proj'' stands for the combination of MSE loss and projection loss. The proposed model uses the combination of MSE, projection and perception losses denoted by ``MSE+perc+proj''. Each image is a maximum intensity projection (MIP) image of a patient's 3D output.}
\label{fig:sup_vis_loss}
\end{figure}

We consider the ablation study on adding segmentation loss $\mathcal{L}_{\textrm{seg}}$ to the CycleGAN loss for the translation module. Fig. \ref{fig:sup_vis_seg} provides a qualitative comparison between the outputs with and without $\mathcal{L}_{\textrm{seg}}$. Table \ref{cyclegan_ablation} presents a quantitative analysis comparing these two cases. While the quantitative difference may appear small, the visual outputs demonstrate substantial differences, revealing that the model utilizing $\mathcal{L}_{\textrm{seg}}$ captures finer dental details compared to the counterpart without $\mathcal{L}_{\textrm{seg}}$.
We observe that the model with $\mathcal{L}_{\textrm{seg}}$ focuses on reconstructing the teeth region, while the one without $\mathcal{L}_{\textrm{seg}}$ perhaps is better in reconstructing other regions, judging from the similar quantitative performance shown in Table \ref{cyclegan_ablation}.
Our observations underscore the effectiveness of the segmentation loss in enhancing the reconstruction of teeth, which is of clinical importance.

\subsubsection{Generation Module} \label{refinement_ablation}
We combined three types of loss functions in the generation module: MSE loss, projection loss and perceptual loss. We conduct an ablation study on the combination of loss functions. The quantitative results are presented in Table \ref{perc_ablation}. The result shows that the combination of three loss functions yields the best result. A visual comparison is also shown in Fig. \ref{fig:sup_vis_loss}. The images show that the inclusion of perceptual loss and projection loss indeed enhances the quality of 3D outputs.

\begin{table*}[h!]
\begin{center}
\setlength\tabcolsep{9pt}
\begin{tabular}{|l|c|c|c|c|}
\hline
Method & PSNR (dB) & SSIM (\%) & Dice (\%) & LPIPS\\
\hline
MSE & 18.23$\pm$0.23 & 77.68$\pm$0.40 & 62.28$\pm$1.49 & 0.36$\pm$0.015\\
MSE+proj & 19.61$\pm$0.21 & 77.68$\pm$0.48 & 64.79$\pm$1.41 & 0.37$\pm$0.015\\
MSE+perc+proj (proposed)& \textbf{19.72}$\pm$\textbf{0.21} & \textbf{78.58}$\pm$\textbf{0.35} & \textbf{67.00}$\pm$\textbf{1.11} & \textbf{0.30}$\pm$\textbf{0.006}\\
\hline
\end{tabular}
\end{center}
\caption{Ablation study on the combination of loss functions in the generation module. ``MSE'' stands for the MSE loss only, ``MSE+proj'' stands for the combination of MSE loss and projection loss. The proposed model uses the combination of MSE, projection and perception losses denoted by ``MSE+perc+proj''.}
\label{perc_ablation}
\end{table*}

\subsection{Model and Implementation Details} \label{implementation}
\subsubsection{Model Configuration}
\paragraph{Translation Module.}

\begin{table}[h]
\caption{Detailed model configuration of translation module.}
\begin{center}
\setlength\tabcolsep{4.5pt}
\begin{tabular}{|l|c|}
\hline
UNet hidden dimension size & 64, 128, 256, 512\\ 
\hline
Segmentation loss weight & $\lambda$=10\\
\hline
Learning rate & $2\times 10^{-4}$\\
\hline
Training epochs & 100\\
\hline
\end{tabular}
\end{center}
\end{table}

The CycleGAN generator in our study uses a UNet architecture consisting of four layers, with feature dimensions of 64, 128, 256, and 512. The loss weight $\lambda$ for segmentation loss is set to 10. The model was trained using a learning rate of $2\times 10^{-4}$ for 100 epochs. For SynA and SynB, we used the same learning rate and the number of training epochs.

\paragraph{Generation Module.}

\begin{table}[h]
\caption{Detailed model configuration of generation module.}
\begin{center}
\setlength\tabcolsep{3pt}
\fontsize{9pt}{9pt}\selectfont
\def\arraystretch{1.3}%
\begin{tabular}{|l|c|}
\hline
\# of layers (Image encoder) & 8\\ 
\hline
Hidden dimension (Image encoder) & 64, 128, 256, 512\\
\hline
Output dimension (Image encoder) & 128\\
\hline
\# of freq. of positional encoder $\gamma$ & 7\\
\hline
\# of layers (MLP) & 8\\ 
\hline
Hidden dimension (MLP) & 128\\
\hline
Model for encoder-decoder & 3D UNet\\
\hline
\# of layers (3D UNet)& 8\\ 
\hline
Hidden dimension (3D UNet)& 64, 128, 256, 512\\ 
\hline
$\lambda_1$, $\lambda_2$ for the loss function& 10, 1\\ 
\hline
Input/output shape & $128 \times 256 \times 256$\\
\hline
\end{tabular}
\end{center}
\end{table}

The input PX image is encoded using a UNet model. The model used four layers of convolutional operations and four layers of up-convolutional operations. The feature size for each convolutional layer was set as 64, 128, 256, and 512, respectively. Moreover, the output dimension of the encoder UNet was fixed at 128 in order to match the size of the hidden dimension of the MLP model.

%Additionally, the positional encoded vector $\gamma(\bm{x})$ played a role as another input to the model.
The positional encoding $\gamma(\cdot)$ is applied to the sampled location. Following the same encoder design in NeRF \cite{mildenhall2020nerf}, $\gamma$ uses sinusoids with increasing frequencies given by $\gamma(p) = (\sin(2^0p), \cos(2^0p), ..., \sin(2^{L-1}p), \cos(2^{L-1}p))$, where $L$ is set to 7 in our study. For $\bm{x}=(x_1,x_2,x_3)$, $\gamma$ is applied to each component $p=x_1, x_2, x_3$. Consequently, a total of 42 size vectors were obtained. These vectors were then processed through a fully connected (FC) layer, resulting in an output vector of size 128.

The encoded output from the UNet is passed to another FC layer, which produced an output vector of size 128. These two 128-sized vectors were added together and used as the final input to MLP.

For the MLP model, we adopted the MLP layer from the NeRF architecture. The model consists of eight layers, including a skip connection at the 4th layer. The hidden dimension was set to 128.

For the encoder-decoder model following the MLP layer, we use a 3D UNet architecture consisting of four levels with feature dimensions of 64, 128, 256, and 512. $\lambda_1$ and $\lambda_2$ in Eq. \ref{eq:nebla_loss} are set to 10 and 1 respectively. The input and output dimensions of the refinement module were set to $128 \times 256 \times 256$ which is the output size of generation module.

\subsubsection{Implementation Details}

\begin{table}[h!]
\caption{Implementation details of our framework and Baselines.}
\begin{center}
\setlength\tabcolsep{4pt}
\fontsize{8pt}{8pt}\selectfont
\def\arraystretch{1.3}%
\begin{tabular}{|l|c|}
\hline
Threshold for visual results& 0.2\\ 
\hline
Optimizer & Adam\\
\hline
Learning rate & $10^{-4}$\\
\hline
Learning rate (X2CT-GAN) & $2\times 10^{-4}$\\
\hline
Batch size & 1 for image and 32768 for rays\\
\hline
Training epochs & 300\\
\hline
Usage of early stopping method & Yes\\
\hline
GPU for experiments & NVIDIA A100 and NVIDIA A6000\\ 
\hline
\end{tabular}
\end{center}
\end{table}

In the training of the generation and refinement modules, Adam optimizer with a learning rate of $10^{-4}$ was used. Both modules were trained with a batch size of 1 for the input PX image, where each image has dimension  $128 \times 256$. Thus, the number of rays for each image is given by $128 \times 256 = 32768$. 
In our settings, the whole batch of 32768 rays is used in each training step.

\subsection{Additional Visualization Results} \label{more_visual}

In this section, we present visualizations of additional cases, showcasing the performance and versatility of our method. These outputs demonstrate the accuracy and high-quality results achieved by our method across different PX images.

\begin{figure*}[h]
\centering
\includegraphics[width=1.0\linewidth]{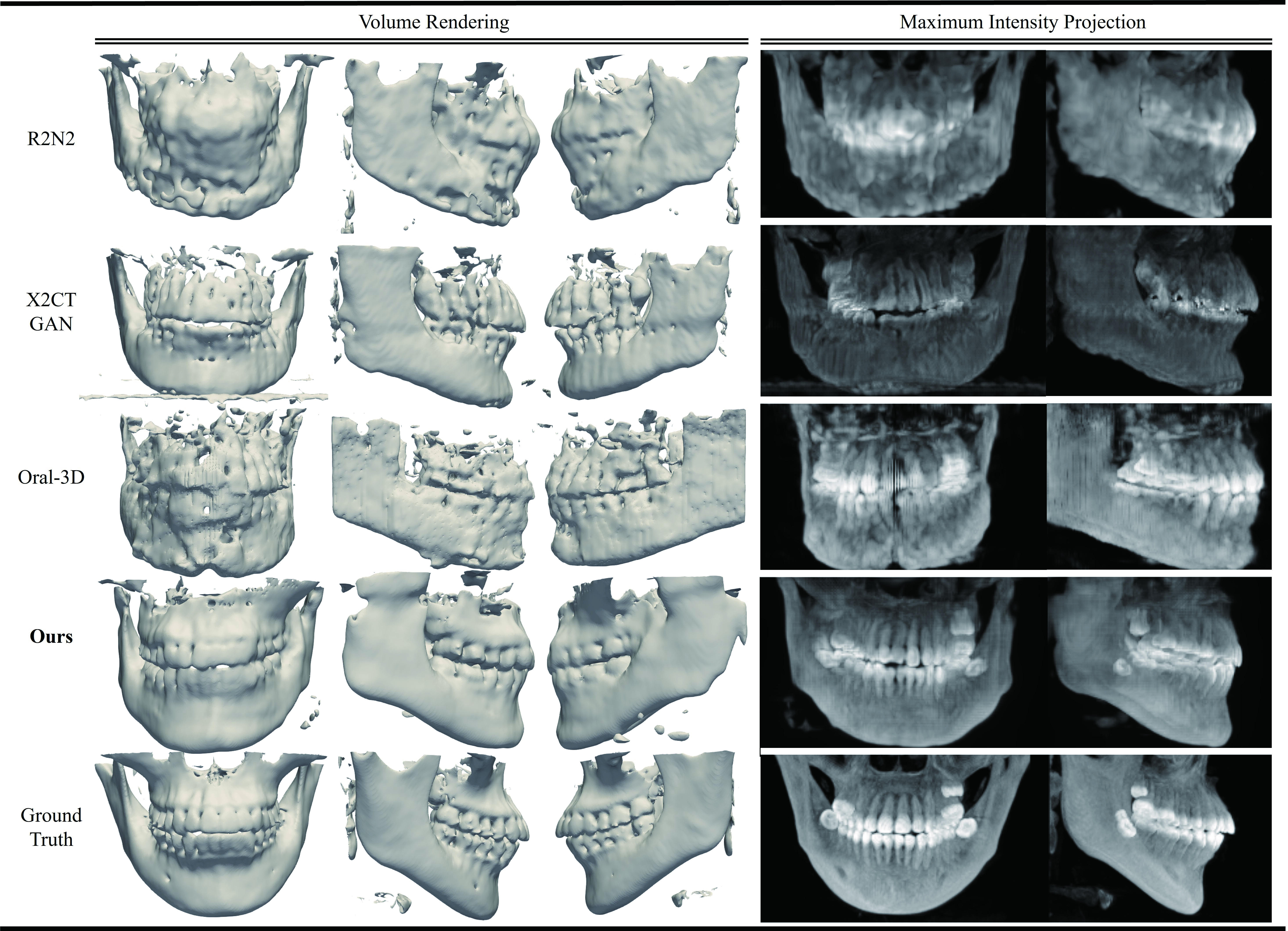}
   \caption{Qualitative comparison of 3D reconstruction results from real PX image (Case A).}
\label{fig:sup_vis_real_1}
\end{figure*} 

\begin{figure*}[h]
\centering
\includegraphics[width=1.0\linewidth]{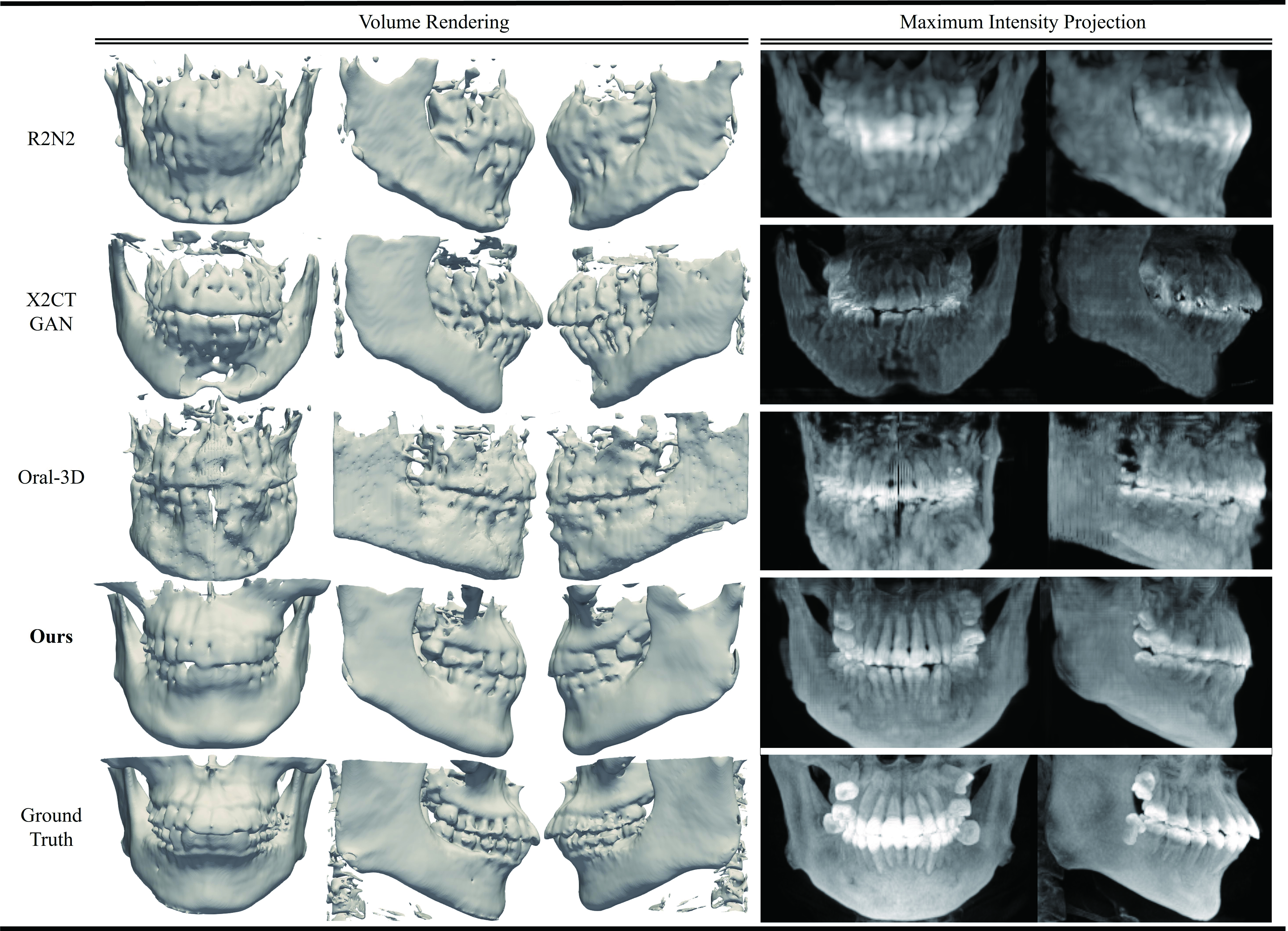}
   \caption{Qualitative comparison of 3D reconstruction results from real PX image (Case B).}
\label{fig:sup_vis_real_2}
\end{figure*} 

\begin{figure*}[h]
\centering
\includegraphics[width=1.0\linewidth]{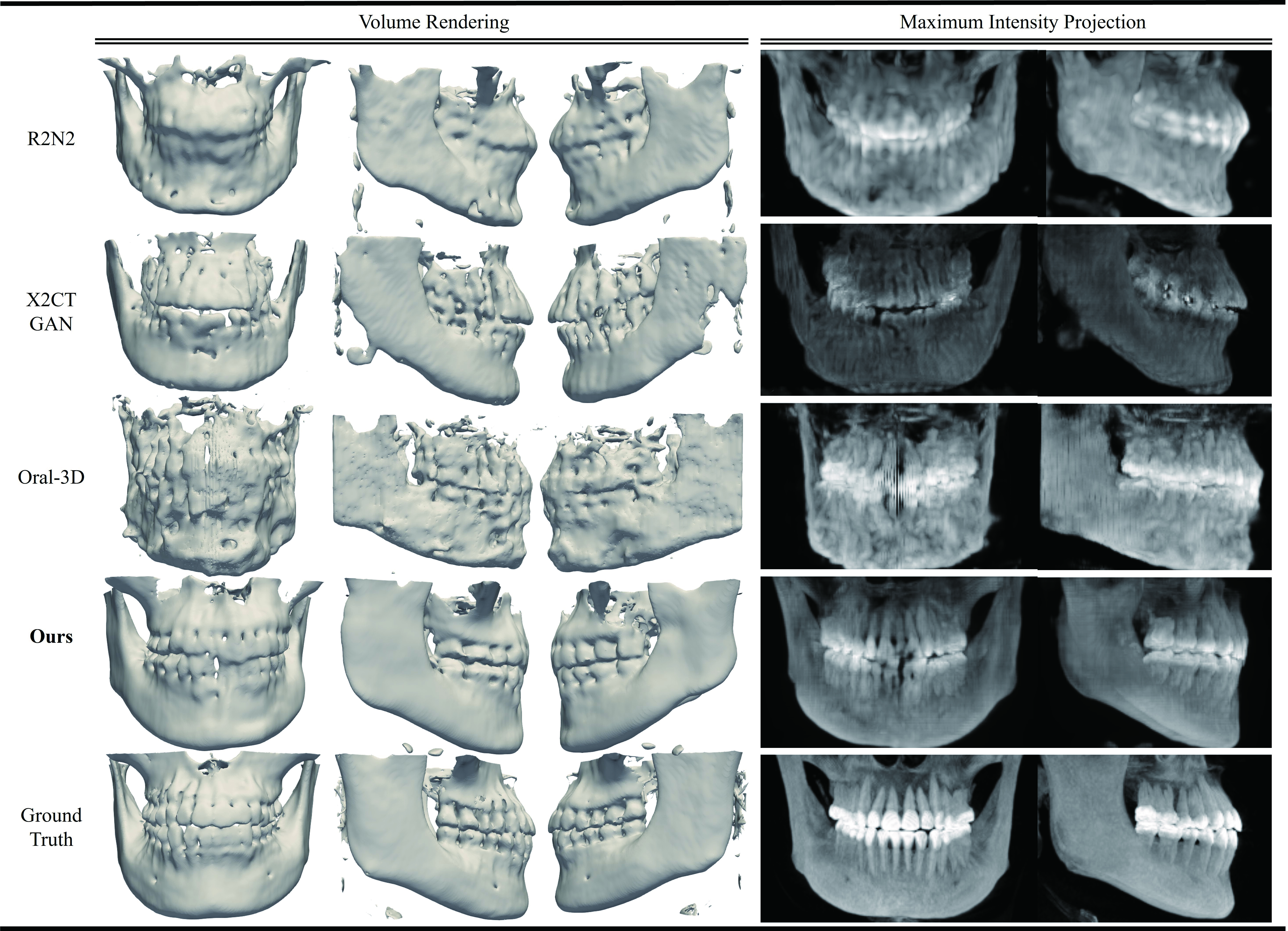}
   \caption{Qualitative comparison of 3D reconstruction results from real PX image (Case C).}
\label{fig:sup_vis_real_3}
\end{figure*} 

\begin{figure*}[h]
\centering
\includegraphics[width=1.0\linewidth]{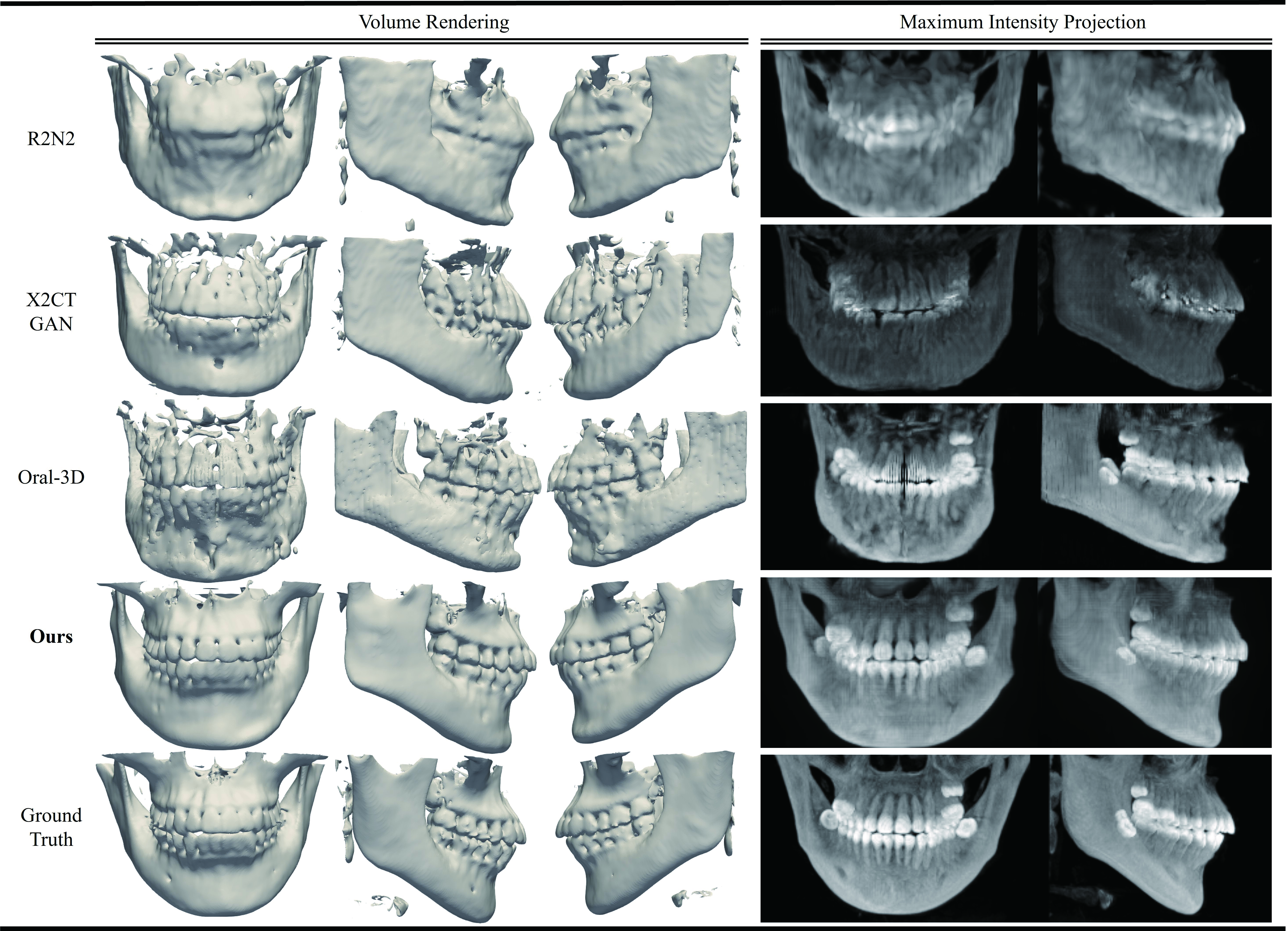}
   \caption{Qualitative comparison of 3D reconstruction results from SimPX image (Case D).}
\label{fig:sup_vis_syn_1}
\end{figure*} 

\begin{figure*}[h]
\centering
\includegraphics[width=1.0\linewidth]{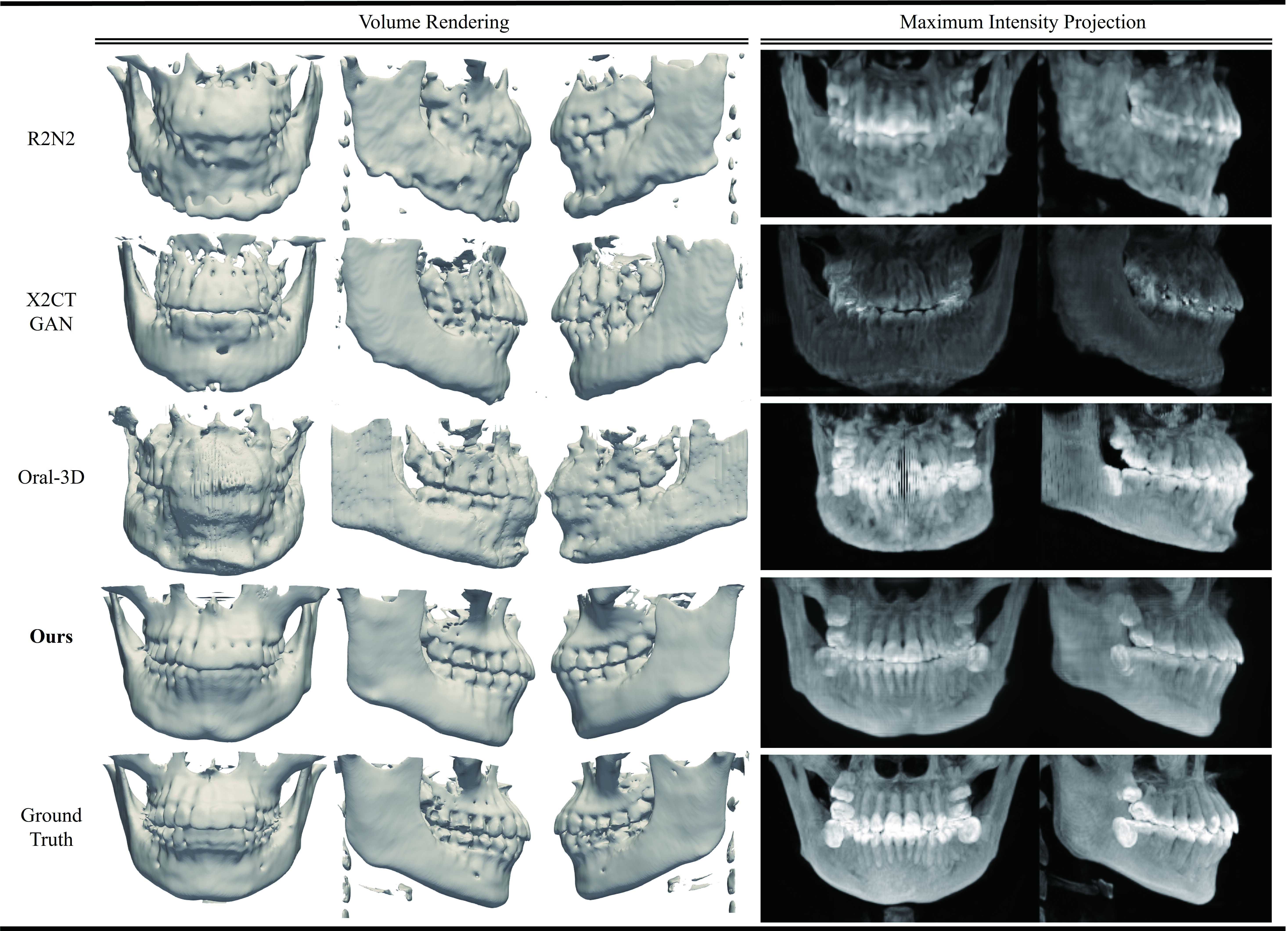}
   \caption{Qualitative comparison of 3D reconstruction results from SimPX image (Case E).}
\label{fig:sup_vis_syn_2} 
\end{figure*} 

\begin{figure*}[h]
\centering
\includegraphics[width=1.0\linewidth]{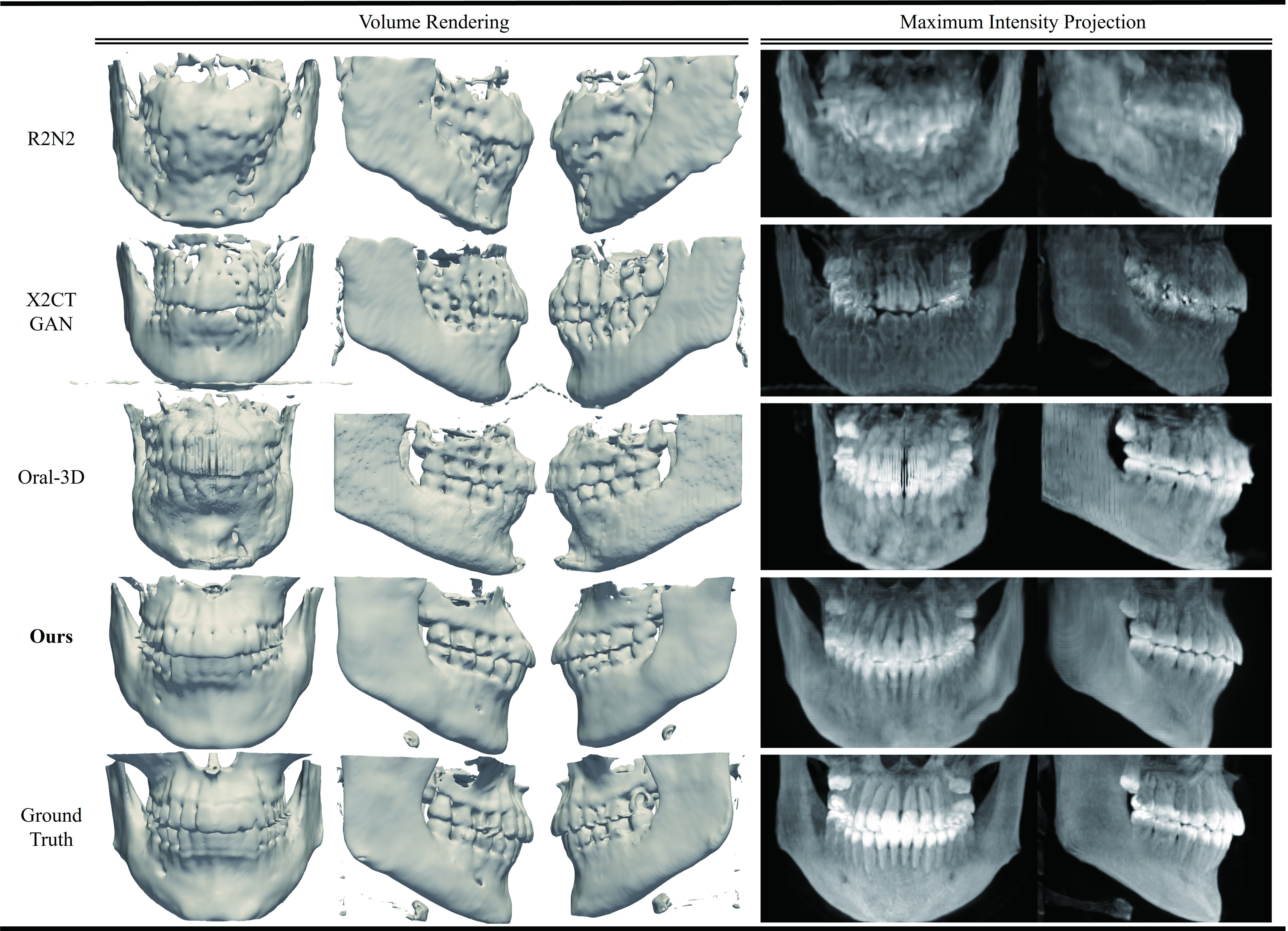}
   \caption{Qualitative comparison of 3D reconstruction results from SimPX image (Case F).}
\label{fig:sup_vis_syn_3}
\end{figure*}
\end{document}